\documentclass{aa}  

\usepackage{graphicx}
\usepackage{color}
\usepackage{amssymb}
\usepackage[varg]{txfonts}
\begin{document}

\title{Multi-instrument analysis of 67P/Churyumov-Gerasimenko coma particles: COPS-GIADA data fusion}

\author{B. Pestoni\inst{1} \and K. Altwegg\inst{1} \and V. Della Corte\inst{2} \and N. Hänni\inst{1} \and A. Longobardo\inst{2} \and D. R. Müller\inst{1} \and A. Rotundi\inst{2,3} \and M. Rubin\inst{1} \and S. F. Wampfler\inst{4}}

\institute{Physics Institute, Space Research \& Planetary Sciences, University of Bern, Sidlerstrasse 5, 3012 Bern, Switzerland\\ \email{boris.pestoni@unibe.ch}
\and
IAPS, INAF-IAPS, via Fosso del Cavaliere 100, 00133 Rome, Italy
\and
Dipartimento di Scienze e Tecnologie, Università degli Studi di Napoli Parthenope, CDN IC4, 80143 Naples, Italy
\and
Center for Space and Habitability, University of Bern, Gesellschaftsstrasse 6, 3012 Bern, Switzerland}

\date{Received XXX / Accepted YYY}

 
\abstract
{The European Space Agency's Rosetta mission to comet 67P/Churyumov-Gerasimenko has offered scientists the opportunity to study a comet in unprecedented detail. Four instruments of the Rosetta orbiter, namely, the Micro-Imaging Dust Analysis System (MIDAS), the Grain Impact Analyzer and Dust Accumulator (GIADA), the COmetary Secondary Ion Mass Analyser (COSIMA), and the Rosetta Orbiter Spectrometer for Ion and Neutral Analysis (ROSINA) have provided information on cometary dust particles. Cross-instrument comparisons are crucial to characterize cometary dust particles beyond the capabilities of individual sensors, as they are sensitive to different dust components.}
{We present the first comparison between detections of the ROSINA COmet Pressure Sensor (COPS) and GIADA. These two instruments are complementary as the former is sensitive solely to volatiles of icy particles, while the latter measured the dust particle as a whole, including refractories and condensed (semi)volatiles. Our goal is to correlate the particles detected by COPS  and GIADA and to assess whether they belong to a common group.}
{We statistically analyzed the in situ data of COPS and GIADA by calculating Pearson correlation coefficients.}
{Among the several types of particles detected by GIADA, we find that COPS particles are significantly correlated solely with GIADA fluffy agglomerates (Pearson correlation coefficient of 0.55 and p-value of $4.6\cdot10^{-3}$). This suggests that fluffy particles are composed of both refractories and volatiles. COPS volatile volumes, which may be represented by equivalent spheres with a diameter in the range between 0.06 $\mu$m and 0.8 $\mu$m, are similar to the sizes of the fractal particle’s subunits identified by MIDAS (i.e., 0.05-0.18 $\mu$m).}
{}


\titlerunning{Multi-instrument analysis of 67P/Churyumov-Gerasimenko coma particles: COPS-GIADA data fusion}
\maketitle


\section{Introduction}\label{sec:introduction}
The Jupiter-family comet 67P/Churyumov-Gerasimenko (hereafter, 67P) was established as the target of the European Space Agency's Rosetta mission. After more than ten years of travel, Rosetta rendezvoused with 67P in early August 2014. In the following two years, Rosetta escorted the comet from a heliocentric distance of 3.6 au to perihelion at 1.24 au and then out to a heliocentric distance of 3.8 au. The mission ended with a soft-landing of the orbiter on the comet's surface at the end of September 2016. 

The 11 instruments of the Rosetta orbiter \citep[][]{Vallat_2017} and the 10 instruments of the lander module Philae \citep[][]{Boehnhardt_2017} have allowed for the most in-depth studies of a comet's nucleus and surrounding coma to date. 

Some of the Rosetta orbiter instruments were specifically developed to investigate dust particles ejected from the nucleus of 67P. The Micro-Imaging Dust Analysis System \citep[MIDAS,][]{Riedler_2007} is an atomic force microscope with the capabilities to resolve micron-sized dust agglomerates constituted by grains with sizes as small as 100 nanometres \citep{Mannel_2019}. Particles detected by MIDAS are, in fact, fragments of larger particles (hundreds of microns to mm-sized) that have directly been ejected from the comet surface \citep[][]{Longobardo_2022}. The Grain Impact Analyzer and Dust Accumulator \citep[GIADA,][]{Della_Corte_2014} was dedicated to measure the speed, momentum, and cumulative mass of the dust particles in 67P's coma. GIADA accumulated the mass of collected nanogram dust particles using micro balances \citep{Della_Corte_2019} and detected impacting millimeter and submillimeter dust particles \citep{Rotundi_2015,Fulle_2015}. A procedure to trace these dust particles to their source regions on the nucleus of 67P was implemented \citep{Longobardo_2020}. The COmetary Secondary Ion Mass Analyser \citep[COSIMA,][]{kissel_2007} was another instrument which probed dust particles. COSIMA collected and acquired images and analyzed tens of thousands of 67P dust particles by secondary ion mass spectrometry \citep{Bardyn_2017}.

Although it was not designed for such a purpose, it was found that the COmet Pressure Sensor \citep[COPS,][]{Balsiger_2007} can also be used to study 67P dust particles' volatile components \citep{Pestoni_2021_RG,Pestoni_2021_NG}. In fact, there are abrupt increases seen in the gas density measured by the instrument when the volatile content of cometary dust particles sublimated in the vicinity or within the instrument. COPS can only investigate the volatile component of dust. In contrast, MIDAS and COSIMA observed only refractories due to the extended time between collection and analysis. GIADA sensed the whole particle, made-up of refractories and condensed (semi)volatiles. The distinction between (semi)volatiles and refractories lies in their different condensation temperatures and since the above-listed instruments detected distinct components of dust particles, the COPS results are complementary to the ones obtained by the other three instruments. 

The purpose of this work is to merge COPS and GIADA data and investigate possible correlations between the two instrument datasets. The results of GIADA have already been compared to those of MIDAS in \citet{Longobardo_2022}. The comparison between the two datasets evidenced that micron particles detected by MIDAS are produced from fragmentation of mm-sized particles (as those retrieved by GIADA) due to dust impacts on the MIDAS funnel and/or target. COPS uncovered icy particles, that is, dust particles composed of a condensed volatile component with a probable addition of a refractory component \citep[][]{Pestoni_2021_RG}: since refractories are invisible to COPS, it is possible that some icy particles (if not all) only consist of condensed volatiles. GIADA collected submicrometer- to micrometer-sized particles \citep{Della_Corte_2019} and larger dust particles which can be classified into two distinct groups according to their structure. The two groups are fragments of fluffy agglomerates with sizes in the range from 0.2~mm to 2.5~mm and compact particles ranging in sizes from 0.03~mm to 1~mm \citep{Rotundi_2015,Fulle_2015}. Pristine fluffy agglomerates are also called parent particles.

Each type of particle gathered by GIADA may contain a condensed (semi)volatile component. This means that according to the GIADA classification \citep[][]{Longobardo_2019}, COPS icy particles may be included in submicrometer- to micrometer-sized particles, parent particles, fragments of fluffy agglomerates, or compact particles.


\section{Instruments}\label{sec:methods}
\subsection{COPS}\label{subsec:cops_description}
COPS is one of the three sensors of the Rosetta Orbiter Spectrometer for Ion and Neutral Analysis \citep[ROSINA,][]{Balsiger_2007}. The other two are mass spectrometers: the Reflectron Time-of-Flight mass spectrometer (RTOF) and the Double Focusing Mass Spectrometer (DFMS), which have been employed to determine the elemental, molecular, and isotopic composition and abundances in 67P \citep[e.g.,][]{Rubin_2019,Altwegg_2020,Mueller_2022,Haenni_2022}.

COPS consists of two gauges: the nude gauge (hereafter, NG) measures the total neutral gas density and the ram gauge (hereafter, RG) measures the ram pressure. The ram pressure is proportional to the 67P gas flux. Both gauges operate according to the extractor-type ionization gauge principle \citep[][]{Redhead_1966}. First, neutral atoms and molecules enter the ionization region and are ionized either by a hot filament (in the NG) or by a microtip field emitter (in the RG). Unless they have sufficient kinetic energy \citep[see][]{Pestoni_2021_NG}, charged atoms and molecules are prevented from entering the ionization region. Ionized atoms and molecules are then accelerated towards a base plate and focused to the collector by a reflector. Finally, the collector current is measured by a sensitive electrometer. This current can then be related to the density within the ionization volume. Proportionality constants are known from calibration experiments performed in the laboratory with a flight spare of the COPS instrument \citep[][]{Graf_2004}. The measurement interval of COPS is usually one minute, but sometimes as small as two seconds. The recorded value is a running average of several seconds.

In addition to the method of ionization, the RG and NG also differ by their design and pointing direction. The RG has a spherical equilibrium cavity on top of the ionization region and points by default towards the comet’s nucleus (S/C z axis). The NG has no physical barriers precluding the sublimating volatiles from escaping the ionization region and sticks out on the side of the Rosetta spacecraft (S/C -y axis). The NG has a large field of view of more than $2\pi$ steradians which included the comet most of the time but also many other directions. It is for these differences in design that the features generated by the sublimation of the volatile component of icy particles in the NG data are much more frequent and have a shorter duration than those in the RG data.

In the absence of icy particles, the two gauges measure a nominal density background generated by gaseous coma species. The nominal density background of the NG is lower and smoother than the one of the RG. Therefore, it is possible to identify many more features generated by the sublimation of the volatile content of 67P dust particles in the NG data than in the RG data \citep[$\sim$67000 vs 73 features,][]{Pestoni_2021_RG,Pestoni_2021_NG}. The difference between the number of features in the two gauges is accentuated by the different cross section (748~mm$^2$ vs.\ 27~mm$^2$) and by the fact that the NG operated for many more days than the RG. In fact, the NG performed measurements almost for the entire mission, whereas the RG was operational only for 319 days out of 791. 

\citet{Pestoni_2021_RG} analyzed in depth 25 features in the RG data and found three different time constants, interpreted as different volatile compositions. No further constraints on the type of volatiles could be derived.

\subsection{GIADA}\label{subsec:giada_description}
GIADA is made up of three subsystems: the MicroBalance System (hereafter, MBS), the Grain Detection System (hereafter, GDS), and the Impact Sensor \citep[hereafter, IS,][]{Della_Corte_2014}. The MBS is used to investigate the dust mass fluency, whereas the GDS and IS are arranged in a cascade to derive information, such as the cross section and the mass of the detected particles.

The MBS is a set of five Quartz Crystal Microbalances (QCMs), each having a field of view of 40$^\circ$, located on the upper plate of GIADA. Overall, QCM5 is the only microbalance that points to the S/C z axis and, hence, nominally to the comet \citep{Della_Corte_2019}. The change in the resonance frequency of the QCMs is proportional to the mass accumulated on top of them \citep{Palomba_2002}. Throughout the mission, the QCMs recorded every five minutes the mass of the accumulated submicrometer- to micrometer-sized dust particles.

The GDS consists of a laser curtain and several photodiode sensors. When a particle crosses the laser curtain, the light is first scattered or reflected and finally detected by the photodiode sensors. The amplitude of the signal measured by the photodiodes is related to the particle geometrical cross section \citep[][]{Mazzotta_2002}. The IS is composed of an aluminum square plate and an array of five piezoelectric sensors. When a particle hits the plate, bending waves are produced and detected by the sensors. From the amplitude of the bending waves it is then possible to determine the particle momentum \citep[][]{Esposito_2002}. A timer is first started when a particle crosses the GDS laser curtain and then stopped when the particle impacts the IS plate. Since the distance between the laser curtain and the plate is known, the particle's velocity can be obtained. Finally, having information on both momentum and velocity, the mass of the dust particle is derived. Dust particles can be detected by the GDS subsystem (GDS-only detections), by the IS subsystem (IS-only detections), and by both the GDS and IS subsystems (GDS-IS detections). Dust modeling \citep[][]{Fulle_2015} and calibration experiments with cometary dust analogues \citep[][]{Ferrari_2014,Della_Corte_2016} allowed us to relate these detection combinations to a dust particle's physical properties \citep[][]{Longobardo_2019}: (1) IS-only and GDS-IS detections are ascribable to compact particles; (2) isolated GDS-only detections are attributable to compact particles having a sideways trajectory that is not compatible with an impact on the IS; and (3) numerous GDS-only detections grouped in space and time are associated with fluffy fragments. Fluffy fragments have a momentum too slow to trigger the IS and are the outcome of the fragmentation, induced by the spacecraft potential, of a single pristine fluffy agglomerate. The latter is identified as parent particle.

Like the NG, GIADA has been operational for almost for the entire mission. GIADA found 2110 compact particles (out of which 284 are isolated GDS-only detections) and 3159 fluffy fragments which arose as a result of the fragmentation of 277 parent particles. QCM5 reported an accumulated mass of 83.5 $\mu$g \citep{Della_Corte_2019}. 


\section{Results}\label{sec:results}
\subsection{Detections analysis}\label{subsec:det_analysis}
We look for periods showing an abundance of particles detected by both COPS and GIADA. In particular, we investigated whether the so-called outbursts leave a particular signature on these two instruments and whether there are times when COPS and GIADA (nearly) simultaneously detect large numbers of particles. Outbursts are transient events characterized by a sudden and brief increase in dust ejection from spatially restricted regions of the comet. Several instruments on board the Rosetta orbiter detected outbursts \citep[e.g.,][and references therein]{Vincent_2016,Rinaldi_2018,Noonan_2021}. It is not possible to find a clear signature associated with these outbursts in both GIADA and COPS. One explanation is that the vast majority of the outbursts were detected near the nucleus and near 90$^\circ$ phase angle by the remote sensing instruments of Rosetta, whereas COPS and GIADA are in-situ instruments that only observe the environment immediately surrounding the spacecraft. Moreover, there is also a time delay in the observations corresponding to the travel time of the dust from the nucleus to Rosetta. 

\begin{figure*}
\centering
\includegraphics[width=0.83\textwidth]{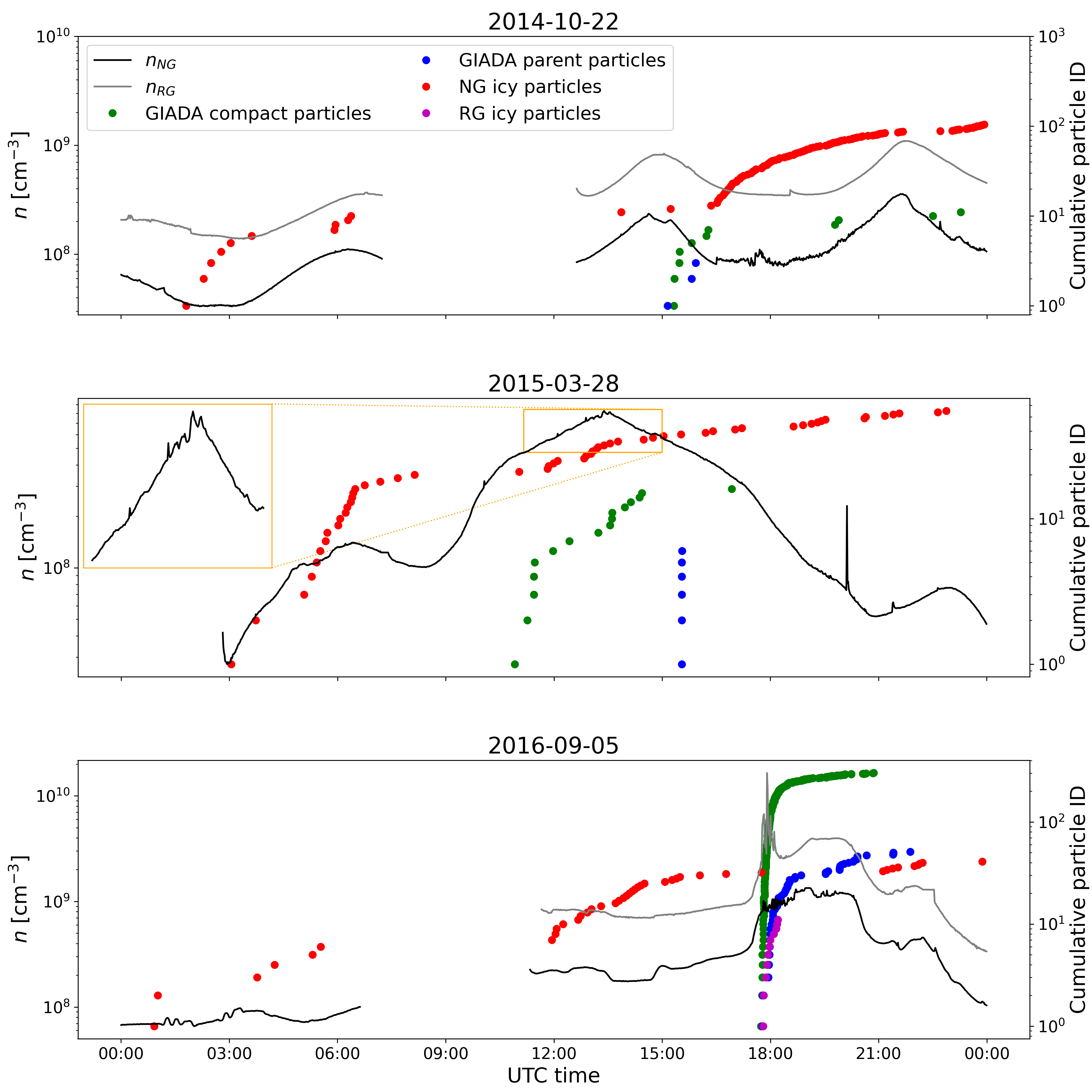}
\caption{Comparison of COPS and GIADA measurements during 22 October 2014, 28 March 2015, and 5 September 2016. The plots display the gas density measured by the two COPS gauges (black curve for the NG, gray curve for the RG), the compact and parent particles detected by GIADA (green and blue dots, respectively), and the icy particles detected by the NG (red dots) and by the RG (magenta dots). Between 7:25 AM UTC and 12:15 PM UTC on 22 October 2014, and between 6:40 AM UTC and 11:00 AM UTC on 5 September 2016, there were orbital correction maneuvers during which COPS did not acquire any data. Large spikes in the NG density data, such as the one at 8:07 PM UTC on 28 March 2015, are associated with thruster operations \citep[][]{Pestoni_2021_NG}. There are also spikes in the RG data which happened following offset measurements \citep[][]{Pestoni_2021_RG}. The most evident one is at 6:33 PM UTC on 22 October 2014.
}
\label{fig:comparison_three_days}
\end{figure*}

Figure~\ref{fig:comparison_three_days} shows COPS and GIADA data from 22 October 2014, 28 March 2015, and 5 September 2016. MBS data are excluded from the plots due to the low statistics. These selected days contain the three periods when the concurrence of the detections of the two instruments is most evident. Other days with similar patterns were noted on 15 March 2015, 1 August 2015, 19 February 2016, 28 May 2016, and 3 July 2016. 

On 22 October 2014, between 3:08 PM UTC and 4:16 PM UTC, GIADA detected six compact particles and fragments of three parent particles. The first compact particle and the fragments of the first parent particle are found by GIADA at the same time as the NG detects an icy particle. Over the next six hours, a couple more compact particles are measured by GIADA, whereas the NG observes ninety icy particles and has a detection rate much higher than the one in the previous and following hours. The NG density pattern concomitant with the particle detections is noisy. Analyzing the full NG dataset, it is found that this is a characteristic that occurs when a large amount of icy particles are detected by COPS. Thus, we can infer that a large amount of dust reached the spacecraft and we conclude that these GIADA and NG detections belong to the same dust event.

On 28 March 2015, a close flyby at the comet was carried out by the spacecraft. Between 9 AM UTC and 6 PM UTC, the NG density pattern (i.e., a bump that lasts almost nine hours) revealed an excess of gas coming from 67P due to the change in cometocentric distance. As can be seen from the zoomed in panel, we identify the noise connected to an excess of dust reaching the spacecraft. During this time period, both the NG and GIADA detect particles. In contrast to the dust event on 22 October 2014, no increase in the NG detection rate is observed. While on 22 October 2014, GIADA simultaneously observed compact particles and fragments of parent particles, on 28 March 2015, these particle groups were detected at separate times. This can be explained by considering that the cometocentric distance in March 2015 is almost double that of October 2014: compact and parent particles are probably ejected together from the nucleus of 67P and then separated in the coma due to their different speed \citep[][]{Longobardo_2020,Longobardo_2022}.

On 5 September 2016, between 5:30 PM UTC and 8:30 PM UTC, DFMS \citep[][]{Altwegg_2017}, GIADA \citep{Della_Corte_2019}, and COPS \citep[][]{Pestoni_2021_RG} reported a huge increase in dust activity. This is compatible with outburst material directly impacting the spacecraft. At the beginning of the outburst, COPS and GIADA detect dust particles nearly at the same time. This is the only situation of the mission where there is a simultaneous detection of the fragments of a parent particle, a compact particle, and two icy particles (one in the RG, one in the NG). During the outburst, the NG observes a single icy particle and the RG detects twelve icy particles between 5:45 PM UTC and 6:12 PM UTC. The absence of detections in the second part of the outburst can be explained by the saturation of the two COPS gauges. In fact, it is impossible to discern features attributable to the sublimation of the volatile content of the icy particles, as they merge in the very high density background. From the noisy pattern of NG density, we can infer that a large amount of icy dust has impacted in the vicinity of COPS throughout the full duration of the outburst. GIADA does not suffer from saturation problems and measures compact particles and fragments of parent particles over the full outburst. Immediately following the end of the outburst, GIADA finds fragments of three parent particles and the NG detects seven icy particles. These particles may have been ejected from the nucleus during the outburst and then taken longer than the others to reach the spacecraft.

\subsection{Correlation analysis}\label{subsec:Corr_analysis}
\begin{figure*}
\centering
\includegraphics[width=\textwidth]{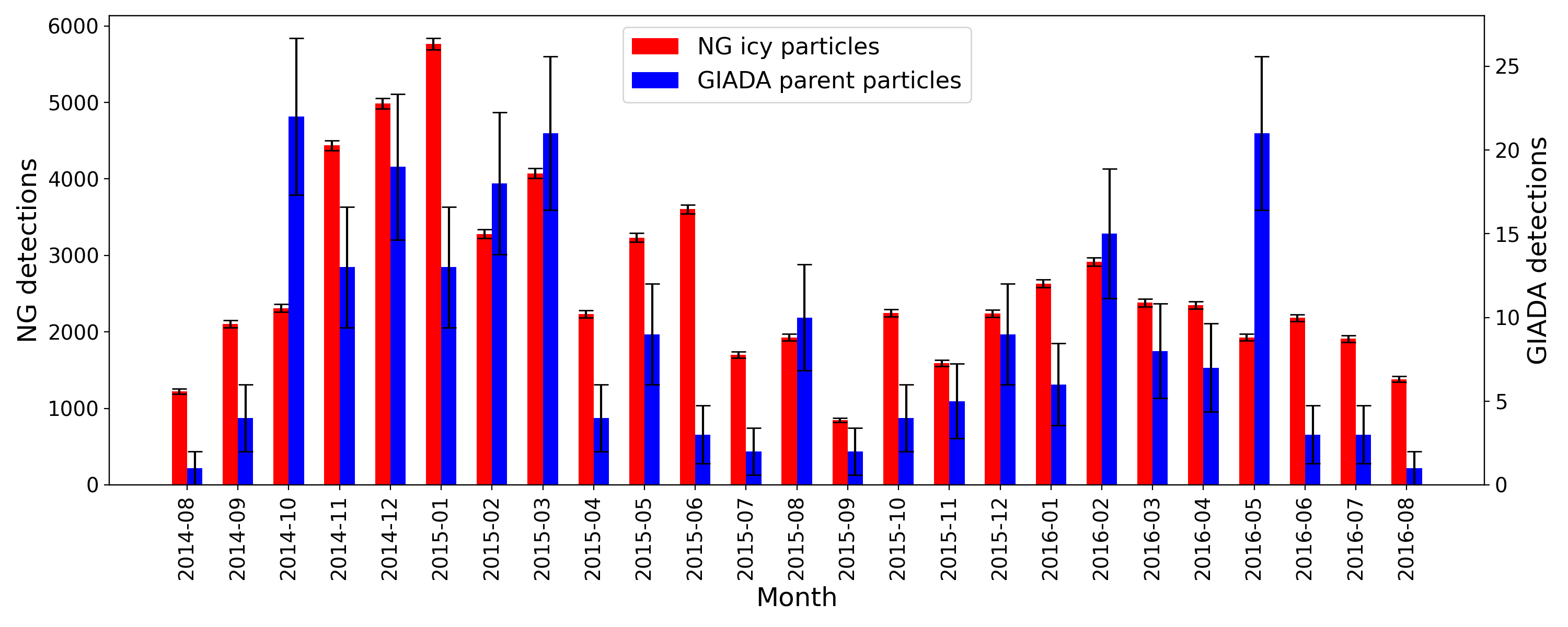}
\caption{Monthly detections of NG icy particles (red bars) and of GIADA parent particles (blue bars).}
\label{fig:NG_icy_vs_GIADA_parent}
\end{figure*}

To search for possible correlations in time between COPS and GIADA, a data selection process was performed. Firstly, September 2016 is excluded from the analysis, as 52 out of 277 parent particles detected by GIADA are from the 5 September 2016 outburst (see Fig.~\ref{fig:comparison_three_days}). Secondly, RG detections are excluded from the investigation because this gauge was turned off for many days of the mission and count rates were anyways negligible compared to the NG (see Sect.~\ref{subsec:cops_description}). This means that the data to be compared are: (1) NG icy particles vs. GIADA compact particles; (2) NG icy particles vs. GIADA parent particles; (3) NG icy particles vs. GIADA fluffy fragments; and (4) NG icy particles vs. GIADA nanogram dust particles. For each of these data pairs, the detections per month were analyzed and a Pearson correlation coefficient was calculated. Together with the Pearson correlation coefficient, we calculated the p-value, a parameter that is useful in determining the significance of the correlation: if the p-value is less than 0.05, then the correlation is statistically significant. The obtained values are listed in Table~\ref{tab:correlations}.
\begin{table}
\centering
\caption{Correlations between COPS particles and GIADA particles.}
\begin{tabular}{p{3.7cm}p{2.0cm}p{1.2cm}}
\hline\hline
Data to correlate with NG icy particles & Pearson correlation coefficient & p-value \\
\hline
GIADA compact particles & 0.05 & 0.80 \\
GIADA parent particles & 0.55 & $4.6\cdot10^{-3}$ \\
GIADA fluffy fragments & 0.37 & 0.07 \\
GIADA nanogram dust particles & -0.06 & 0.78 \\
\hline
\end{tabular}
\label{tab:correlations}
\end{table}
It can be seen that NG icy particles are best correlated with GIADA parent particles, while none of the other correlations are significant. The correlation between NG icy particles and GIADA parent particles is moderate. Nevertheless, from the plot in Fig.~\ref{fig:NG_icy_vs_GIADA_parent}, the correlation can be visually appreciated, as the trend of the two instruments is very similar. This trend is largely governed by the cometocentric distance, which was generally large when 67P was close to the perihelion and small when 67P was at greater heliocentric distance.

\begin{figure*}
\centering
\includegraphics[width=\textwidth]{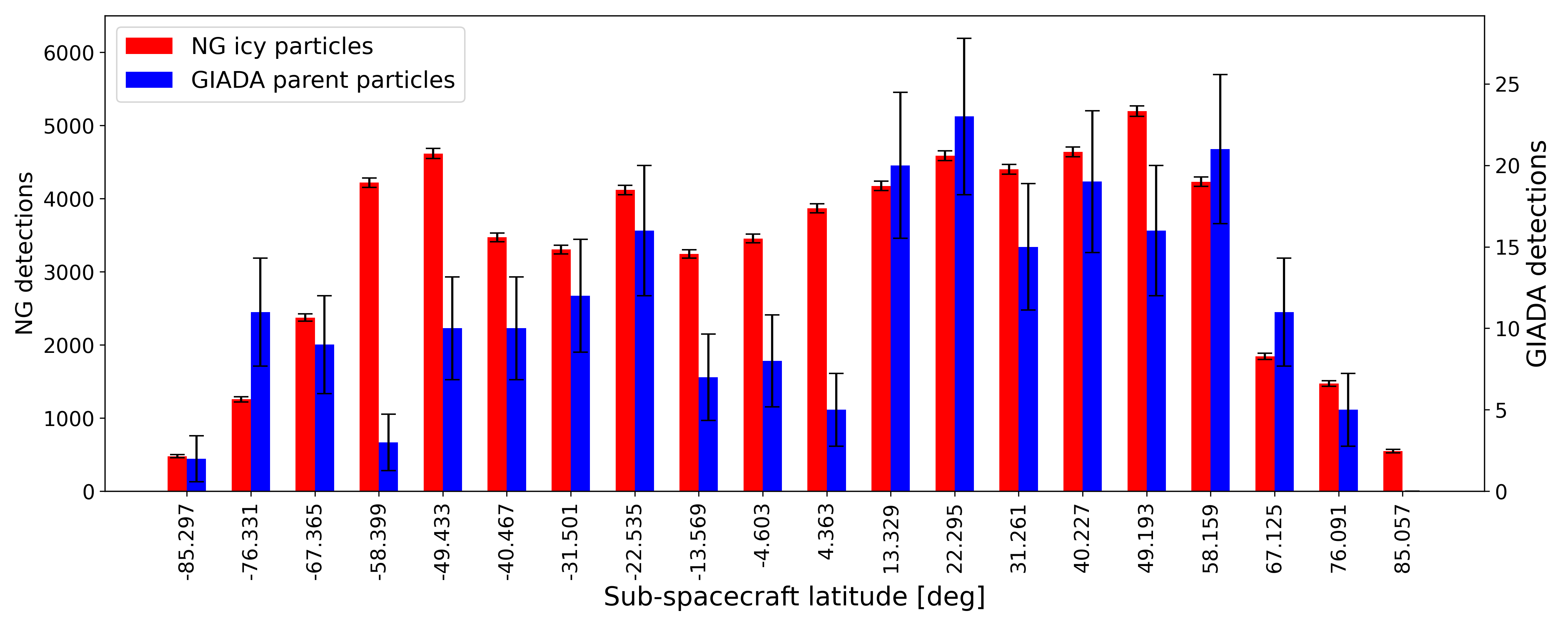}
\caption{Number of NG icy particles detections (red bars) and of GIADA parent particles detections (blue bars) as a function of sub-spacecraft latitude. On the x-axis, the central values of bins with a width of 8.966 deg are shown.}
\label{fig:NG_icy_vs_GIADA_parent_latitude}
\end{figure*}

The relationship between NG icy particles and GIADA parent particles is also supported by the analysis of detections as a function of subspacecraft latitude (see Fig.~\ref{fig:NG_icy_vs_GIADA_parent_latitude}). In fact, NG icy particles and GIADA parent particles exhibit a Pearson correlation coefficient of 0.65 (p-value: $1.9\cdot10^{-3}$), whereas the other GIADA particle classes have a much lower correlation.

The missing correlation with compact particles may be a clue to infer the structure of these particles. The possibilities include: (1) volatiles in compact particles sublimate either when lifted from the nucleus or on the way to the spacecraft, possibly due to better heat conduction within the grain compared to fluffy particles; (2) compact particles do not include any volatiles; (3) the amount of volatiles and composition in compact particles varies strongly and hence the correlation to the volatile content measured by the NG is less pronounced; and (4) volatiles are embedded within refractories and sublimate at such low rates that they end up blending with the background coma gas. Unlike fluffy particles, compact particles do not fragment, so the volatile fraction is not released. 

Since parent particle are fragmented by the Rosetta spacecraft potential \citep[][]{Fulle_2015}, the NG may not have observed whole parent particles. Consequently, NG detections are the outcome of one or more fluffy fragments reaching the instrument within the COPS time resolution of one minute. Moreover, no evident shower of fluffy fragments is visible in the COPS data. This is probably due to the COPS time resolution. 

\citet{Pestoni_2021_RG,Pestoni_2021_NG} estimated volumes of the volatile part of the icy particles detected by COPS by assuming a density of 1 g cm$^{-3}$ and that each COPS event corresponds to only one icy particle. These volumes are represented by equivalent spheres with a diameter in the range between 0.06 $\mu$m and 0.8 $\mu$m. As already stated in \citet{Pestoni_2021_NG}, it should be cautioned that the diameters derived from NG data may be lower limits of the amount of volatiles of the dust particles. In fact, it is possible that only a fraction of the sublimated volatiles may have entered the ionization region. Furthermore, a portion of volatiles may be invisible to COPS if they remain trapped within refractories. The diameters derived from NG data are comparable to the ones derived from RG data. The RG has a closed design and the flux of outgoing volatiles from the opening of the spherical equilibrium cavity is low \citep[][]{Pestoni_2021_RG}. This legitimates the assumption that the estimated volumes are representative of the amount of volatiles of the fluffy fragments identified by COPS. 

To estimate the number of icy fragments that generated the features in the COPS data, for each event, the diameter of the equivalent sphere is compared to the sizes of the subunits of the fractal particle detected by MIDAS \citep[52-183 nm,][]{Mannel_2019}. From this comparison, it naturally follows that the COPS gauges measured an icy equivalent of 1 to 15 fluffy subunits per impact event within its time resolution of one measurement per minute.

\citet{Davidsson_2021} modeled the transfer of material from one hemisphere to the other of 67P and concluded that a significant amount of water ice is kept during this process even in cm-sized particles. These particles are much larger than those detected by COPS. Nevertheless, COPS would also measure species of lower volatility compared to water, such as the numerous salts and organics found by DFMS \citep[e.g.,][]{Altwegg_2020,Altwegg_2022,Haenni_2022}. The volatility of many of these identified species is low enough to survive, at least in parts, the travel between 67P's surface and the spacecraft even if the situation is complicated due to varying cometocentric distance of the Rosetta spacecraft coupled with heliocentric distance and dust sizes, velocities, composition, structure, and shapes.

Even considering that fluffy particles have a fractal dimension of $\sim2$ \citep[][]{Guttler_et_al_2019}, the volumes of volatiles during a single impact detected by either of the two COPS gauges are much smaller than the size of parent particles measured by GIADA \citep[0.2-2.5~mm,][]{Rotundi_2015,Fulle_2015}. This observation can be explained in two ways. Either parent particles are almost totally dehydrated, possibly due to their fractal structure receiving more sunlight, or the NG may also see particles outgassing away from the ionization zone due to its large field of view. Hence, only a fraction of the volatiles are seen. These possible explanations require in-depth investigations both via simulations and laboratory experiments with cometary analogue materials.


\section{Conclusions}\label{sec:conclusion}
In this paper, COPS and GIADA detections are merged to gain new information on 67P dust particles. First, we found very few periods with an excess of particle detections by both COPS and GIADA. This is probably due to the complex interplay of geometric parameters (heliocentric and cometocentric distances, sub-spacecraft latitude, phase angle, and nadir angle) and cometary activity throughout the mission. Furthermore, apart from the one on 5 September 2016, outbursts detected by other instruments of the mission \citep[e.g.,][and references therein]{Vincent_2016,Rinaldi_2018,Noonan_2021} cannot be seen in the measurements of both COPS and GIADA data. This is due to the fact that the instruments that detected outbursts see the latter preferentially looking near 90$^\circ$ phase angle. The in situ instruments detect outbursts only when the Rosetta spacecraft is flying over their footpoint. 

The most significant result of the COPS-GIADA data fusion is the identification of a correlation between NG icy particles and GIADA parent particles, that is, fluffy agglomerates which were fragmented by the Rosetta spacecraft potential \citep[][]{Fulle_2015} before reaching the instruments. This is a clear indication that fluffy particles contain both a refractory and a volatile component. This is relevant since our knowledge on structure and composition of icy grains in the coma of a comet is limited. Outgassing from icy grains have been observed (and modeled) for comets Schwassmann–Wachmann \citep[][]{Fougere_2012} and Hartley 2 \citep[][]{Ahearn_2011,Fougere_2013,Protopapa_2014}. However, the detailed structure and mixture of these grains are unknown.

Taking the subunits size of the fractal particles detected by MIDAS as reference \citep[][]{Mannel_2019}, we conclude that the two gauges of COPS sensed between 1 to 15 icy subunits of fluffy particles arriving within one interval of measurement. Due to the low time resolution, COPS cannot temporally resolve single fragments of fluffy particles, especially if parent particles fragment near the spacecraft. The volume of volatiles estimated based on COPS density data is very small compared to the size of the whole fluffy particles. Future studies are needed to investigate the causes and implications of this finding.


\begin{acknowledgements}
Work at the University of Bern was funded by the State of Bern, the Swiss National Science Foundation (200020\textunderscore182418 \& 200020\textunderscore207312), and the European Space Agency through the Rosetta data fusion: Dust and gas coma modelling (5001018690) grant. SFW acknowledges the financial support of the SNSF Eccellenza Professorial Fellowship (PCEFP2\textunderscore181150). Rosetta is an European Space Agency (ESA) mission with contributions from its member states and NASA. We thank herewith the work of all the technicians, engineers, and scientists of the whole ESA Rosetta team. We would also thank the International Space Science Institute (ISSI) team ``Characterization Of Cometary Activity Of 67P/Churyumov-Gerasimenko Comet'' for their essential contributions.
All ROSINA and GIADA flight data have been released to the Planetary Science Archive (PSA) of ESA and to the Planetary Data System (PDS) archive of NASA.

\end{acknowledgements}

\bibliographystyle{aa} 
\bibliography{Pestoni_et_al_cops_giada_data_fusion} 

\end{document}